\documentclass[%
reprint,
groupedaddress,
 amsmath,amssymb,
 aps,
prl
]{revtex4-2}

\usepackage{graphicx}
\usepackage{dcolumn}
\usepackage{bm}
\usepackage{hyperref}
\hypersetup{
    colorlinks=true,
    linkcolor=blue,
    citecolor=blue,     
    urlcolor=blue,
}

\usepackage{mathtools}
\allowdisplaybreaks
\newcommand{\ii}{\mathrm{i}}
\newcommand{\ee}{\mathrm{e}}

\begin{document}

\preprint{APS/123-QED}

\title{Quantum fluctuation-induced first-order breaking of time-reversal symmetry in unconventional superconductors}

\author{Yin Shi}
 \email{yin.shi@iphy.ac.cn}
 \affiliation{%
 Beijing National Laboratory for Condensed Matter Physics and Institute of Physics, Chinese Academy of Sciences, Beijing 100190, China
}%



\date{\today}

\begin{abstract}
    Spontaneous time-reversal symmetry breaking in superconductors with competing non-degenerate pairing channels is an exotic quantum phase transition that could give rise to robust topological superconductivity and unusual magnetism.
    It is proposed mostly in two-dimensional systems and is signaled by a nonzero relative phase between the two superconducting order parameters, hence it should particularly be prone to order-parameter phase fluctuations.
    Nevertheless, the existing understanding of it is still at the mean-field level.
    Here, we illustrate the non-negligible effects of the phase fluctuations on such quantum phase transitions using the hole-doped square-lattice $t$--$J$ model as an example.
    We derive the phase fluctuation-corrected free energy and show that under the quantum phase fluctuations, the time-reversal asymmetric $s + \ii d$ phase region splits off a dome featuring a first-order border with the $d$ phase, indicating the possibility of a phase separation into the time-reversal symmetric and asymmetric phases.
    The phase fluctuations also narrow the range of the $s + \ii d$ phase considerably.
    We further discuss the implications of our findings for recent experiments on disorder-induced first-order quantum breakdown of superconductivity and promising high-temperature topological superconductivity in twisted cuprate Josephson junctions.
\end{abstract}
\maketitle


\paragraph{Introduction---}
Superconductivity with spontaneously broken time-reversal symmetry $\mathcal{T}$ is an exotic state of matter that has attracted significant attention.
This kind of superconducting state itself can be topologically nontrivial~\cite{read00paired,moore91nonabelions,sauls94the}, which holds the promise to realize robust quantum computers protected from quantum decoherence.
This is often the ground state when two degenerate pairing channels coexist and linearly combine, e.g., the degenerate $p_x$ and $p_y$ channels in a square lattice yield a $p_x + \ii p_y$ ground state~\cite{read00paired,rice95sr2ruo4,ishida98spin} and the degenerate $d_{x^2-y^2}$ and $d_{xy}$ channels in a triangular lattice yield a $d_{x^2-y^2} + \ii d_{xy}$ ground state~\cite{pathak10possible,nandkishore12chiral,ganesh14theoretical,liu13d,wu23pair}, but their conclusive experimental evidence is still lacking or suffering from controversial observations~\cite{kallin09is,pustogow19constraints}.

An alternative and seemingly more promising route for realizing $\mathcal{T}$-broken superconductivity is to tune the competition between two non-degenerate pairing channels in established superconductors to go through a quantum phase transition.
For example, with regard to the non-degenerate $d_{x^2-y^2}$ and $d_{xy}$ channels in cuprate superconductors, an experiment~\cite{krishana97plateaus} has hinted at a magnetically induced quantum phase transition from the $d_{x^2-y^2}$ phase into the $d_{x^2-y^2} + \ii d_{xy}$ phase~\cite{laughlin98magnetic}.
Moreover, a $\mathcal{T}$-broken topological $d_{x^2-y^2}+e^{\ii \vartheta}d_{xy}$ state was recently predicted to be robustly stabilized at high temperatures in twisted cuprate Josephson junctions via the competition between the native $d_{x^2-y^2}$ channel and a Josephson coupling-induced subdominant $d_{xy}$ channel~\cite{yang18josephson,can21high,volkov23current,volkov23magic}, and its $\mathcal{T}$-broken Josephson ground state was further observed in an experiment~\cite{zhao23time}.
Here $\vartheta$ can be changed continuously from $0$ to $\pi$ as the twist angle is changed around $45^\circ$, and hits $\pi/2$, i.e., forming the $d_{x^2-y^2} + \ii d_{xy}$ state, when the twist angle is exactly $45^\circ$~\cite{can21high}.
Therefore, this is also a quantum phase transition but induced by the twisting between the cuprate layers.

The $\mathcal{T}$-broken but topologically trivial $s + \ii d$ state can also emerge in cuprate Josephson junctions via a similar mechanism as described above~\cite{kuboki96proximity,yang18josephson,can21high}, and in homogeneous systems already with competing non-degenerate $d$- and (extended) $s$-wave channels.
A representative example of such homogeneous systems is the square-lattice $t$--$J$ (or Hubbard) model~\cite{li21superconductor,breio22supercurrents,pupim25adatom}, which is regarded as the minimal model of cuprate superconductors.
In this model, hole doping determines the leading pairing channel so that it can induce a quantum phase transition from either the $d$ or extended $s$ ($s^*$) phase into an intermediate $s + \ii d$ phase~\cite{breio22supercurrents}.
This property has recently been used to propose an adatom strategy for engineering nontrivial magnetic orders coexisting with superconductivity~\cite{pupim25adatom}.

As already made clear by their notations, the $\mathcal{T}$-broken superconducting phases arising from competing non-degenerate orders are signaled by a nonzero relative phase between the two order parameters.
Therefore, we anticipate that the quantum phase transitions into these phases, occurring mostly in two-dimensional systems, are particularly vulnerable to dynamical order-parameter phase fluctuations, which are nevertheless ignored in previous theoretical treatments that are all at the mean-field level~\cite{laughlin98magnetic,yang18josephson,can21high,volkov23current,volkov23magic,kuboki96proximity,li21superconductor,breio22supercurrents,pupim25adatom}.
Indeed, the quantum order-parameter phase fluctuations have a dramatic impact on the superconducting ground state of (quasi-) two-dimensional superconductors with disorder and/or (partially) flat bands, e.g., they can greatly suppress the zero-temperature superfluid density~\cite{emery95importance,bozovic16dependence,lee18optical,lee20low,li21superconductor,he21superconducting}.

In this Letter, we illustrate the non-negligible effects of the order-parameter phase fluctuations on the $\mathcal{T}$-breaking quantum phase transition, employing the square-lattice $t$--$J$ model for concreteness.
Using the path integral formalism, we derive the free energy incorporating the long-range order-parameter phase fluctuations coupled to charge density fluctuations in a \emph{self-consistent} manner.
Our calculations show that the quantum phase fluctuations cause the $s + \ii d$ phase region to split off a dome featuring a first-order border with the $d$ phase, implying that there could be a phase separation into the $d$ and $s + \ii d$ parts.
It is also shown that the quantum fluctuations narrow the range of the $s + \ii d$ phase considerably.
The formalism is extendable to other more complex systems.

\paragraph{Model and formalism---}
The square-lattice Hubbard model and its related $t$--$J$ model are regarded as the minimal model for describing superconductivity in cuprates.
Since excited states such as the phase fluctuations and plasmons are relevant, to properly account for them we have to add the long-range Coulomb interaction into the $t$--$J$ model, and the resulting Hamiltonian reads
\begin{equation}
    H = -\sum_{\bm{rr}'\sigma} \tilde{t}_{\bm{r}-\bm{r}'} c_{\bm{r}\sigma}^\dagger c_{\bm{r}'\sigma} - \frac{J}{2}\sum_{\langle \bm{rr}' \rangle} b_{\bm{rr}'}^\dagger b_{\bm{rr}'} + \sum_{\bm{q}} \frac{V_{\bm{q}}}{2N} \rho_{\bm{q}}^\dagger \rho_{\bm{q}}. \label{eq:H}
\end{equation}
Here $c_{\bm{r}\sigma}$ annihilates an electron with spin $\sigma$ on the site at $\bm{r}$.
$b_{\bm{r}\bm{r}'}=c_{\bm{r}\downarrow} c_{\bm{r}'\uparrow} - c_{\bm{r}\uparrow} c_{\bm{r}'\downarrow}$ annihilates a singlet pair.
$\rho_{\bm{q}}=\sum_{\bm{r}} \ee^{-\ii \bm{q}\cdot\bm{r}} n_{\bm{r}}$ is the Fourier component of the electron density $n_{\bm{r}}=\sum_\sigma c_{\bm{r}\sigma}^\dagger c_{\bm{r}\sigma}$.
$\tilde{t}_{\bm{r}-\bm{r}'}=t_{\bm{r}-\bm{r}'}+\mu\delta_{\bm{r}\bm{r}'}$ where $t_{\bm{r}-\bm{r}'}$ is the hopping amplitude and $\mu$ is the chemical potential.
$J>0$ is the nearest-neighbor antiferromagnetic exchange interaction.
$N$ is the number of sites.
$V_{\bm{q}}=e^2/(2\epsilon_b \epsilon_0 a q)$ is the Coulomb potential in the two-dimensional momentum space where $e$ is the elementary charge, $\epsilon_b$ is the background dielectric constant, $\epsilon_0$ is the vacuum permittivity, and $a$ is the lattice constant.
In this Letter we make both $\bm{r}$ and $\bm{q}$ dimensionless by dividing the position vector and multiplying the momentum with $a$. 
Note that in Eq.~(\ref{eq:H}) we have expressed the short-range exchange interaction (the $J$ term) of the $t$--$J$ model in terms of the singlet annihilation and creation operators.

The action after the Hubbard-Stratonovich transformation to decouple the interactions is
\begin{align}
    S = \int_0^{1/T} \mathrm{d} \tau \Biggl\{& \sum_{\bm{rr}'} \psi_{\bm{r}}^\dagger(\tau) \bigl[ \delta_{\bm{rr}'}\bigl(\partial_\tau + \ii \phi_{\bm{r}}(\tau)\tau_3\bigr) - \tilde{t}_{\bm{r}-\bm{r}'}\tau_3 \nonumber \\
    &+ \Delta_{\bm{rr}'}(\tau) \tau_+ + \Delta_{\bm{rr}'}^*(\tau) \tau_- \bigr] \psi_{\bm{r}'}(\tau) \nonumber \\
    &+ \frac{2}{J}\sum_{\langle \bm{rr}' \rangle} |\Delta_{\bm{rr}'}(\tau)|^2 + \sum_{\bm{q}} \frac{|\phi_{\bm{q}}(\tau)|^2}{2 N V_{\bm{q}}} \Biggr\}, \label{eq:S}
\end{align}
where $\Delta_{\bm{rr}'}(\tau)$ and $\phi_{\bm{q}}(\tau)$ are the auxiliary fields for $b_{\bm{rr}'}$ and $\rho_{\bm{q}}$, respectively.
$\phi_{\bm{r}}(\tau)$ is $\phi_{\bm{q}}(\tau)$ transformed to real space.
$\psi_{\bm{r}}(\tau)=(c_{\bm{r}\uparrow}(\tau), c_{\bm{r}\downarrow}^\dagger(\tau))^{\mathrm{T}}$ is the Nambu spinor.
$\tau_+=(\tau_1+\ii\tau_2)/2$, $\tau_-=(\tau_1-\ii\tau_2)/2$, and $\tau_3$ are the Pauli matrices in the Nambu space.
$T$ is temperature in energy unit.

Since we are considering quantum phase transitions at low temperatures, it is sufficient to take into account only the low energy excitations, i.e., the gapless long-range fluctuations of the phase of $\Delta_{\bm{rr}'}(\tau)$ (Nambu-Goldstone mode).
The Berezinskii–Kosterlitz–Thouless vortex--antivortex pair excitations have a finite core-energy cost so that they will be exponentially suppressed at low temperatures, and similarly for the gapped fluctuations of the amplitude of $\Delta_{\bm{rr}'}(\tau)$.
Therefore, we can demand the gauge transformations $\psi_{\bm{r}}(\tau)\rightarrow \psi_{\bm{r}}(\tau) \ee^{\ii\theta_{\bm{r}}(\tau)\tau_3}$ and $\Delta_{\bm{rr}'}(\tau)\rightarrow \Delta_{\bm{r}-\bm{r}'}\ee^{\ii[\theta_{\bm{r}}(\tau)+\theta_{\bm{r}'}(\tau)]}$ where $\Delta_{\bm{r}-\bm{r}'}$ is the saddle-point solution (order parameter), and then expand the action in powers of the small spatiotemporal gradients of $\theta_{\bm{r}}(\tau)$ and $\phi_{\bm{r}}(\tau)$ up to the quadratic order~\cite{paramekanti00effective,sun20collective,yang21theory,yang25efficient,yang25preformed}.
Note that $\Delta_{\bm{r}-\bm{r}'}$ here is allowed to have an arbitrary phase depending only on $\bm{r}-\bm{r}'$, and is nonzero only for $\bm{r}-\bm{r}'=\pm\hat{\bm{x}}, \pm\hat{\bm{y}}$ with $\Delta_{\hat{\bm{x}}}=\Delta_{-\hat{\bm{x}}}\equiv\Delta_x$ and $\Delta_{\hat{\bm{y}}}=\Delta_{-\hat{\bm{y}}}\equiv\Delta_y$ due to $b_{\bm{rr}'}=b_{\bm{r}'\bm{r}}$.
Then after integrating out all the fields in the action, we obtain the free energy per site (the derivation steps are summarized in End Matter)
\begin{align}
    F(\Delta_x,\Delta_y) ={}& -\frac{1}{N} \sum_{\bm{k}} \bigl[ 2T \ln\bigl(1 + \ee^{- E_{\bm{k}}/T}\bigr) + E_{\bm{k}} \bigr] \nonumber \\
    &+ \frac{2}{J}\bigl(|\Delta_x|^2+|\Delta_y|^2
    \bigr) \nonumber \\
    &+ \frac{1}{N} \sum_{q<q_c} \bigl[ 2T \ln\bigl(1 - \ee^{-\Omega_{\bm{q}}/T}\bigr) + \Omega_{\bm{q}} \bigr], \label{eq:F}
\end{align}
where $E_{\bm{k}}=\sqrt{\xi_{\bm{k}}^2+|\Delta_{\bm{k}}|^2}$ and $\Omega_{\bm{q}}$ is the Nambu-Goldstone mode spectrum in the long-wavelength limit
\begin{equation}
    \Omega_{\bm{q}}^2 = \frac{\chi V_{\bm{q}}+1}{\chi}\sum_{\alpha,\beta=x, y} q_\alpha q_\beta D_{\alpha\beta}.
\end{equation}
Here $\xi_{\bm{k}}=-\sum_{\bm{r}}\ee^{-\ii \bm{k}\cdot \bm{r}}\tilde{t}_{\bm{r}}$ is the single-particle energy band, $\Delta_{\bm{k}}=2(\Delta_x \cos k_x + \Delta_y \cos k_y)$ is the Fourier component of $\Delta_{\bm{r}}$, and
\begin{align}
    \chi ={}& \frac{1}{N} \sum_{\bm{k}} \frac{\partial}{\partial \xi_{\bm{k}}} \biggl[ \frac{\xi_{\bm{k}}}{E_{\bm{k}}} \tanh\biggl(\frac{E_{\bm{k}}}{2T}\biggr) \biggr], \\
    D_{\alpha\beta} ={}& \frac{1}{N} \sum_{\bm{k}} \frac{\partial}{\partial E_{\bm{k}}} \biggl[ \frac{1}{E_{\bm{k}}} \tanh\biggl(\frac{E_{\bm{k}}}{2T}\biggr) \biggr] \frac{1}{E_{\bm{k}}} \nonumber \\
    & \times \biggl( \frac{\xi_{\bm{k}}}{2} \frac{\partial |\Delta_{\bm{k}}|^2}{\partial k_\alpha} - |\Delta_{\bm{k}}|^2 \frac{\partial \xi_{\bm{k}}}{\partial k_\alpha} \biggr) \frac{\partial \xi_{\bm{k}}}{\partial k_\beta}
\end{align}
are the compressibility and phase stiffness, respectively.
In Eq.~(\ref{eq:F}), the first two terms constitute the mean-field free energy, whereas the last term is the correction from the order-parameter phase fluctuations, which are valid only within a small cutoff wavevector $q_c\simeq 2|\Delta_x|/v_F$ ($v_F$ is the average Fermi velocity)~\cite{paramekanti00effective,fischer18short}.
Note that $\Omega_{\bm{q}}\sim \sqrt{q}$ for $q\rightarrow 0$ as already established, so the phase fluctuations can be significant while maintaining long-range order at finite temperatures.

The last step is minimizing $F(\Delta_x,\Delta_y)$, which corresponds to a \emph{fully self-consistent} determination of the order parameters in the presence of their phase fluctuations.
Therefore, this will renormalize the compressibility $\chi$ and the phase stiffness $D_{\alpha\beta}$ through the renormalization of $\Delta_x$ and $\Delta_y$, in contrast to the previous theory~\cite{paramekanti00effective}.
We use the nearest-neighbor hopping $t\approx 0.25$~eV, the next-nearest-neighbor hopping $-0.2t$, and the next-next-nearest-neighbor hopping $0.1t$, consistent with the typical values used in low-energy tight-binding models for cuprates~\cite{nicoletti10high,worm24fermi}.
Other hoppings are considered zero.
We use $a \approx 5.4$~\r{A}~\cite{torardi88structures} and $\epsilon_b \approx 4.5$~\cite{levallois16temperature} typical for cuprates and choose a $J=1.2t$.

\paragraph{Phase diagram---}
\begin{figure}
    \centering
    \includegraphics{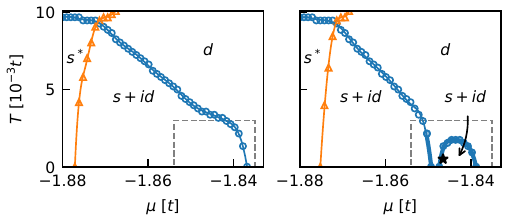}
    \caption{The phase diagrams without (left panel) and with (right panel) the phase fluctuations in the chemical potential--temperature plane. The range of chemical potential $\mu\in (-1.88t, -1.833t)$ corresponds to a range of hole doping $p\in (0.626, 0.640)$ for $T\rightarrow 0$ ($T=10^{-5}t$). The lines are smooth fit to the data points, with the filled circles and thick lines (open circles and thin lines) indicating the first-order (second-order) phase transition. The gray dashed rectangles specify the zoom-in region for Figs.~\ref{fig:gap} and \ref{fig:stiff}. The black star in the right panel marks the state of Fig.~\ref{fig:F}.}
    \label{fig:PD}
\end{figure}

The common phase of $\Delta_x$ and $\Delta_y$ is unimportant and $|\Delta_x|=|\Delta_y|$ due to the symmetry of the square lattice, so the order parameters only have two (instead of four) independent components. 
Therefore, it is convenient and equivalent to rewrite $\Delta_x$ and $\Delta_y$ as $\Delta_x=(\Delta_s+\ii\Delta_d)/2$ and $\Delta_y=(\Delta_s-\ii\Delta_d)/2$ with $\Delta_s$ and $\Delta_d$ being \emph{real} numbers, which correspond to the superconducting gaps in the $s^*$- and $d$-wave channels, respectively.
The gap function is then
\begin{equation}
    \Delta_{\bm{k}}=\Delta_s(\cos k_x+\cos k_y) + \ii \Delta_d (\cos k_x - \cos k_y),
\end{equation}
meaning that simultaneously nonzero $\Delta_s$ and $\Delta_d$ unambiguously signal the $\mathcal{T}$-broken $s + \ii d$ phase, even if the magnitudes of $\Delta_s$ and $\Delta_d$ are not close.

Figure~\ref{fig:PD}, left panel shows the calculated phase diagrams in the chemical potential--temperature plane without the phase fluctuations. 
In the doping range of interest, as the chemical potential is increased, the $s^*$-channel transition temperature (blue line) decreases while the $d$-channel one (orange line) increases, generating the $s + \ii d$ phase~\cite{breio22supercurrents,pupim25adatom}.
This is because the model~(\ref{eq:H}) exhibits a competition between the $s^*$- and $d$-wave superconductivity arising from the change in the Fermi surface shape with doping. 
This doping range corresponds qualitatively to the overdoped regime of cuprates where the $d$-wave superconductivity is about to disappear as the hole doping increases.
The phase transitions from the $s + \ii d$ to either $s^*$ or $d$ phases are second order.

The gap function can also be written as $\Delta_{\bm{k}}=|\Delta_x|(\cos k_x + \ee^{\ii \vartheta}\cos k_y)$, where $\vartheta = -2 \arctan(\Delta_d/\Delta_s)$ is the phase of $\Delta_y$ relative to that of $\Delta_x$ and can vary continuously.
A $\vartheta \neq 0, \pm\pi$ signals the $\mathcal{T}$-broken $s + \ii d$ phase.
It is then obvious that the phase $\vartheta$ and hence the $\mathcal{T}$-breaking phase transition should be strongly influenced by the phase fluctuations.
Indeed, in the presence of the phase fluctuations, the $s + \ii d$ phase region near the quantum critical point bordering the $d$ phase splits off a small dome, which is connected to the $d$ phase by a first-order phase transition (Fig.~\ref{fig:PD}, right panel).
These appear only at low temperatures, which underscores the dramatic effect of the quantum phase fluctuations as compared to the thermal ones.
On the other hand, the $s^*$-to-$s + \ii d$ phase transition maintains second order at all temperatures and its phase boundary is not significantly altered by the phase fluctuations.
The $s + \ii d$ phase region gets narrowed considerably compared to the mean-field case.

\paragraph{First-order quantum phase transition---}
\begin{figure}
    \centering
    \includegraphics{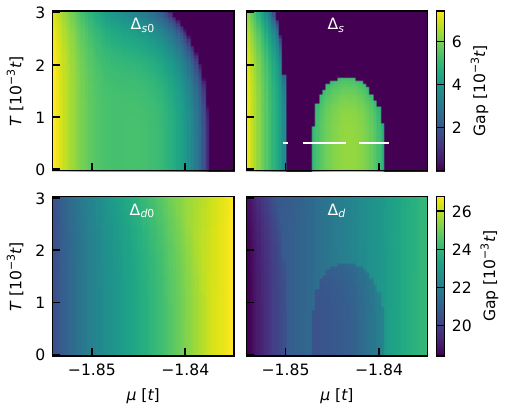}
    \caption{The superconducting gaps without (left column, denoted with subscript $0$) and with (right column) the phase fluctuations as functions of the chemical potential and temperature around the $s + \ii d$ dome. The line segments in the upper right panel indicate the range of existence of a metastable state for $T = 0.0005 t$.}
    \label{fig:gap}
\end{figure}

\begin{figure}
    \centering
    \includegraphics{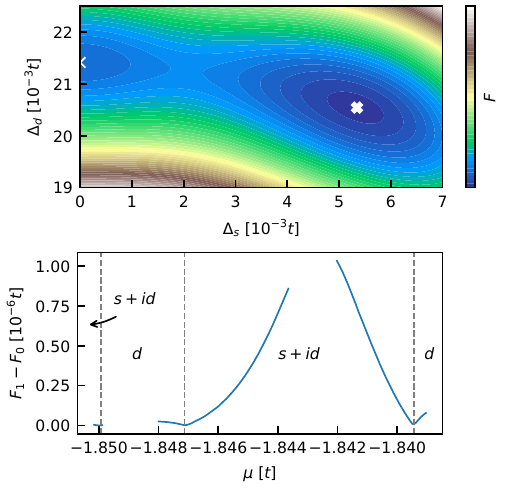}
    \caption{Top: The free energy landscape as functions of the order parameters at $T=0.0005t$ and $\mu=-1.8465t$. This state is marked by the black star in Fig.~\ref{fig:PD}, right panel. The bold and thin crosses mark the stable $s + \ii d$ and metastable $d$ phases, respectively. Note that the free energy is symmetric about $\Delta_s = 0$. Bottom: Difference of the metastable-state energy $F_1$ and stable-state energy $F_0$ as a function of the chemical potential around the $s + \ii d$ dome at $T = 0.0005 t$. Line breaks mean no metastable phase. Gray dashed lines mark the phase boundaries.}
    \label{fig:F}
\end{figure}

To look into the first-order $d$-to-$s + \ii d$ quantum phase transition, in Fig.~\ref{fig:gap} we zoom in on the superconducting gaps around the small $s + \ii d$ dome.
For $T \gtrsim 0.002t$, the mean-field gaps $\Delta_{s0}$ and $\Delta_{d0}$ and the phase-fluctuating gaps $\Delta_s$ and $\Delta_d$ are all changing continuously with the chemical potential and temperature.
But for $T \lesssim 0.002t$, while $\Delta_{s0}$ and $\Delta_{d0}$ are continuous, $\Delta_s$ and $\Delta_d$ clearly exhibit discontinuous jumps, demonstrating the first-order nature of the quantum phase transition.
We further plot the free energy landscape at $T=0.0005t$ and $\mu=-1.8465t$ in Fig.~\ref{fig:F}, top panel.
This particular state is marked in Fig.~\ref{fig:PD}, right panel by a star, showing that it is close to the quantum critical point.
The free energy has two minima: the lower one at finite $\Delta_s$ and $\Delta_d$ corresponds to the stable $s + \ii d$ state, and the higher one at zero $\Delta_s$ and finite $\Delta_d$ corresponds to the metastable $d$ state.
The coexistence of stable and metastable states is an unambiguous signature of first-order phase transitions, and it happens because the phase fluctuations tend to ``mix'' the two states that are distinguished precisely by the relative phase between the competing order parameters.
The metastable state is close to the stable state in energy and appears only around the phase boundary but survives much deeper inside the $s + \ii d$ dome (Fig.~\ref{fig:F}, bottom panel and Fig.~\ref{fig:gap}, upper right panel).

It then follows that for a certain range of hole doping, there could be a spontaneous phase separation into the $d$ and $s + \ii d$ phases.
This can be an interesting state because the phase difference between the $d$ and $s + \ii d$ domains will generate supercurrents leading to spontaneous magnetism~\cite{breio22supercurrents,pupim25adatom} and possible self-organization of the domains.
Since we derived the free energy, it is possible to simulate these mesoscopic phenomena.

From Fig.~\ref{fig:gap} we also see that the phase fluctuations suppress both $\Delta_s$ and $\Delta_d$, except for states inside the $s + \ii d$ dome, in which $\Delta_d$ is suppressed but $\Delta_s$ is enhanced.
Therefore, the ratio $\Delta_s/\Delta_d$ is closer to one than $\Delta_{s0}/\Delta_{d0}$ inside the dome, meaning that the phase fluctuations counterintuitively enhance the $\mathcal{T}$-broken character there.

\begin{figure}
    \centering
    \includegraphics{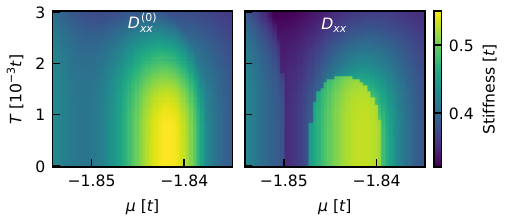}
    \caption{Phase stiffness without [left panel, denoted with superscript $(0)$] and with (right panel) the phase fluctuations as functions of the chemical potential and temperature around the $s + \ii d$ dome.}
    \label{fig:stiff}
\end{figure}

Figure~\ref{fig:stiff} shows the phase stiffness without and with the phase fluctuations.
The stiffness tensor for the square lattice satisfies $D_{\alpha\beta}=D_{xx}\delta_{\alpha\beta}$.
The mean-field stiffness $D_{xx}^{(0)}$ is a continuous function of the chemical potential and temperature, whereas the phase-fluctuating stiffness $D_{xx}$ is discontinuous, reflecting again the first-order nature of the quantum phase transition.
Moreover, the phase fluctuations suppress the phase stiffness even at zero temperature around the $s + \ii d$ dome, i.e., near the quantum critical point.
We find that there is almost no stiffness suppression for chemical potentials far away from the quantum critical point.
Given that our system is clean, this result recognizes the important role of the $\mathcal{T}$-breaking quantum critical point in suppressing the phase stiffness in the ground state.

\paragraph{Discussion---}
The formalism we presented is essentially a \emph{self-consistent} (because we minimized the \emph{total} free energy that includes the phase-fluctuation correction) Gaussian approximation for the phase fluctuations that couple to the long-range charge fluctuations at the level of the random phase approximation.
Because it is a self-consistent approach, this formalism takes into account some higher-order effects of the phase fluctuations beyond the bare Gaussian approximation.
It should be these higher-order effects that modify the order of the $\mathcal{T}$-breaking quantum phase transition.

Increasing the cutoff wavevector within a limit leads to a more profound modification of the phase diagram but retains the qualitative features presented in this work.
However, if $q_c/|\Delta_x|$ is set beyond the limit, e.g., $q_c=4|\Delta_x|/v_F$, we find that the saddle point solution will be unstable at high temperatures well above the superconducting transition temperature~\cite{paramekanti00effective}.

A recent experiment showed that in an amorphous superconductor, increasing disorder quickly widens the temperature window of the pseudogap (phase fluctuating) phase and finally leads to a first-order quantum phase transition from the superconducting phase to a glassy insulator~\cite{charpentier25first}.
This is consistent with the theoretical anticipation that disorder is more influential in promoting the superconducting phase fluctuations and altering the nature of a quantum phase transition compared with a clean system~\cite{charpentier25first}.
In this context, we showed that the inherent phase fluctuations in a clean system can also alter the order of a $\mathcal{T}$-breaking quantum phase transition within superconducting phases, demonstrating the vulnerability of $\mathcal{T}$-broken superconductivity to the order-parameter phase fluctuations.

Finally, the formalism presented in this work can be extended straightforwardly for other systems such as twisted cuprate Josephson junctions promising to host high-temperature topological $d_{x^2-y^2} + \ii d_{xy}$ superconductivity~\cite{can21high,zhao23time}.
In this case, the only extension is to include multiple orbitals because of the enlarged unit cell of the bilayer system.
Due to the similar mathematical structures of the formulations,
although we have not directly calculated for this twist-induced $d_{x^2-y^2} + \ii d_{xy}$ phase, our current findings may still provide insights for interpreting the recent experiment~\cite{zhao23time} on it.
The experiment showed an intricate feature: the reversible Josephson diode effect, which implies breaking $\mathcal{T}$, sharply vanishes for some twist angles within $(40^\circ, 44^\circ)$ (Fig. 4F, G in Ref.~\cite{zhao23time}).
This suggests that the $d_{x^2-y^2} + \ii d_{xy}$ phase does not exist for some twist angles where the mean-field theory predicts to support it~\cite{can21high} and that the associated quantum phase transition is of first order.
It is, however, consistent with our findings of the phase fluctuation-induced splitting of the $\mathcal{T}$-broken phase region and modification of the phase-transition order near the quantum critical point.
It will be a promising direction to use our formalism to directly study chiral superconductivity in twisted cuprate Josephson junctions and to make quantitative comparisons with experimental data.

\begin{acknowledgments}
    I thank Fei Yang, Zi-Xiang Li, and Yi-Ming Wu for useful discussions.
    This work was supported by the start-up grant from the Institute of Physics, Chinese Academy of Sciences.
\end{acknowledgments}


\bibliography{refs}

\appendix

\begin{widetext}
\section{End Matter}
We summarize here the derivation steps for the free energy.
Plugging the gauge transformations $\psi_{\bm{r}}(\tau)\rightarrow \psi_{\bm{r}}(\tau) \ee^{\ii\theta_{\bm{r}}(\tau)\tau_3}$ and $\Delta_{\bm{rr}'}(\tau)\rightarrow \Delta_{\bm{r}-\bm{r}'}\ee^{\ii[\theta_{\bm{r}}(\tau)+\theta_{\bm{r}'}(\tau)]}$ into Eq.~(\ref{eq:S}), we obtain
\begin{equation}
    S = \int_0^{1/T} \mathrm{d}\tau \biggl\{ -\sum_{\bm{rr}'} \psi_{\bm{r}}^\dagger(\tau) \bigl[ \mathcal{G}^{-1}(\bm{r}-\bm{r}', \tau) - \Sigma(\bm{r}, \bm{r}', \tau) \bigr] \psi_{\bm{r}'}(\tau) + \sum_{\bm{q}} \frac{|\phi_{\bm{q}}(\tau)|^2}{2 N V_{\bm{q}}} \biggr\} + \sum_{\langle \bm{rr}' \rangle} \frac{2 |\Delta_{\bm{r}-\bm{r}'}|^2}{TJ},
\end{equation}
where
\begin{align}
    \mathcal{G}^{-1}(\bm{r},\tau) &= -\delta_{\bm{r0}}\partial_\tau + \tilde{t}_{\bm{r}} \tau_3 - \Delta_{\bm{r}} \tau_+ - \Delta_{\bm{r}}^*\tau_-, \\
    \Sigma(\bm{r}, \bm{r}', \tau) &= \Bigl\{ \ii \bigl[ \partial_\tau \theta_{\bm{r}}(\tau) + \phi_{\bm{r}}(\tau) \bigr] \delta_{\bm{rr}'} - t_{\bm{r}-\bm{r}'} \bigl[\ee^{-\ii (\theta_{\bm{r}}(\tau) - \theta_{\bm{r}'}(\tau)) \tau_3} - 1\bigr] \Bigr\} \tau_3,
\end{align}
are the inverse mean-field Green's function and self-energy, respectively.
Invoking the Fourier transformations $\psi_{\bm{r}}(\tau) = N^{-1/2} \sum_{\bm{k}n} \ee^{-\ii (\omega_n \tau - {\bm k} \cdot {\bm r})} \psi_{{\bm k}n}$, $\theta_{\bm{r}}(\tau) = N^{-1/2} \sum_{\bm{k} n} \ee^{-\ii(\nu_n \tau - \bm{k} \cdot \bm{r})} \theta_{\bm{k} n}$, $\phi_{\bm{r}}(\tau) = N^{-1} \sum_{\bm{k} n} \ee^{-\ii(\nu_n \tau - \bm{k} \cdot \bm{r})} \phi_{\bm{k} n}$, and $\Delta_{\bm{r}} = N^{-1} \sum_{\bm{k}} \ee^{\ii \bm{k} \cdot \bm{r}} \Delta_{\bm{k}}$, we get
\begin{align}
    S ={}& -\frac{1}{T}\sum_{\bm{k} \bm{q}} \sum_{mn} \psi_{{\bm k} m}^\dagger \bigl[ \mathcal{G}^{-1}({\bm k}, \ii\omega_n) \delta_{\bm{k} \bm{q}} \delta_{mn} - \Sigma({\bm k}, {\bm q}, \ii\nu_{m - n}) \bigr] \psi_{{\bm q} n} \nonumber \\
    &+ \frac{2}{NTJ} \sum_{\bm{k} \bm{q}} \bigl[ \cos(k_x-q_x) + \cos(k_y-q_y) \bigr] \Delta_{\bm{k}} \Delta_{\bm{q}}^* + \sum_{\bm{q} n} \frac{|\phi_{\bm{q} n}|^2}{2 N T V_{\bm{q}}}, \label{eq:SF}
\end{align}
where $\mathcal{G}^{-1}(\bm{k}, \ii \omega_n) = \ii \omega_n - \xi_{\bm{k}} \tau_3 - \Delta_{\bm{k}} \tau_+ - \Delta_{\bm{k}}^*\tau_-$ and $\Sigma(\bm{k}, \bm{q}, i \nu_n) = (T/N) \int_0^{1/T} d\tau \sum_{\bm{rr}'} \ee^{\ii (\nu_n\tau - \bm{k} \cdot \bm{r} + \bm{q} \cdot \bm{r}')} \Sigma(\bm{r}, \bm{r}', \tau)$.
Here $\omega_n=(2\pi+1)nT$ and $\nu_n=2\pi nT$ with $n$ being an integer are the fermionic and bosonic Matsubara frequencies, respectively.
Integrating out the fermion field $\psi_{\bm{k}n}$ in Eq.~(\ref{eq:SF}), we get the effective action
\begin{equation}
    S_{\mathrm{eff}} = -\mathrm{Tr} \bigl[\ln \bigl( (T\mathcal{G})^{-1} \bigr)\bigr] + \sum_{\gamma = 1}^\infty \frac{1}{\gamma} \mathrm{Tr}\bigl[ (\mathcal{G} \Sigma)^\gamma \bigr] + \frac{2N}{TJ} \bigl( |\Delta_x|^2+|\Delta_y|^2 \bigr) + \sum_{\bm{q} n} \frac{|\phi_{\bm{q} n}|^2}{2 N TV_{\bm{q}}}, \label{eq:Seff}
\end{equation}
where the trace is over the momentum, frequency, and Nambu subspaces.

We then expand $\Sigma(\bm{r},\bm{r}',\tau)$ up to the quadratic order of $\theta_{\bm{r}}(\tau)-\theta_{\bm{r}'}(\tau)$, which is small due to the long wavelength of $\theta_{\bm{r}}(\tau)$ at low temperatures,
\begin{equation}
    \Sigma(\bm{k},\bm{q},\ii\nu_n) \approx \frac{1}{\sqrt{N}} \bigl[ \nu_n \tau_3 + \ii(\xi_{\bm{k}} - \xi_{\bm{q}}) \bigr] \theta_{\bm{k} - \bm{q}, n} - \frac{1}{2 N} \sum_{\bm{k}' m} (\xi_{\bm{k}} + \xi_{\bm{q}} - 2 \xi_{\bm{k} - \bm{k}'}) \tau_3 \theta_{\bm{k}' m} \theta_{\bm{k} - \bm{q} - \bm{k}', n - m} + \frac{\ii}{N} \phi_{\bm{k} - \bm{q}, n} \tau_3.
\end{equation}
We also keep only $\gamma=1,2$ terms in Eq.~(\ref{eq:Seff}) to retain the phase terms up to the quadratic order of $\theta_{\bm{r}}(\tau)-\theta_{\bm{r}'}(\tau)$ in the effective action,
\begin{align}
    \mathrm{Tr}[\mathcal{G}\Sigma] \approx{}& \frac{1}{NT} \sum_{\bm{k} \bm{q}n} \frac{\xi_{\bm{k}} (\xi_{\bm{k}} - \xi_{\bm{k} - \bm{q}})}{E_{\bm{k}}} \tanh\biggl(\frac{E_{\bm{k}}}{2T}\biggr) |\theta_{\bm{q} n}|^2, \label{eq:g1} \\
    \mathrm{Tr}\bigl[(\mathcal{G}\Sigma)^2\bigr] \approx{}& \frac{1}{N} \sum_{\bm{k} \bm{q}mn} \mathrm{Tr}\Bigl[ \mathcal{G}(\bm{k}, \ii\omega_n) \bigl[ \bigl( \nu_m \theta_{\bm{q} m} + N^{-1/2} \ii \phi_{\bm{q} m} \bigr) \tau_3 + \ii (\xi_{\bm{k}} - \xi_{\bm{k} - \bm{q}}) \theta_{\bm{q} m} \bigr] \nonumber \\
    &\times \mathcal{G}(\bm{k} - \bm{q}, \ii\omega_{n - m}) \bigl[ \bigl( -\nu_m \theta_{-\bm{q}, -m} + N^{-1/2} \ii \phi_{-\bm{q}, -m} \bigr) \tau_3 - \ii (\xi_{\bm{k}} - \xi_{\bm{k} - \bm{q}}) \theta_{-\bm{q}, -m} \bigr] \Bigr]. \label{eq:g2}
\end{align}
Since $\theta_{\bm{r}}(\tau)$ is varying slowly in space and time, $\theta_{\bm{q}n}$ would quickly vanish as $\bm{q}$ and $n$ depart from zero.
We can thus further expand Eqs.~(\ref{eq:g1}, \ref{eq:g2}) in powers of $\bm{q}$ and $\nu_n$ up to their quadratic order because they are controlled by $\theta_{\bm{q}n}$, that is, expand $\xi_{\bm{k}} - \xi_{\bm{k} - \bm{q}}$ to the quadratic order of $\bm{q}$ in Eq.~(\ref{eq:g1}) and replace $\mathcal{G}(\bm{k} - \bm{q}, \ii\omega_{n - m})$ with $\mathcal{G}(\bm{k}, \ii\omega_n)$ and expand $\xi_{\bm{k}} - \xi_{\bm{k} - \bm{q}}$ to the linear order of $\bm{q}$ in Eq.~(\ref{eq:g2}).
We obtain
\begin{align}
    \mathrm{Tr}[\mathcal{G} \Sigma] &\approx \frac{1}{2T} \sum_{\bm{q} n} \sum_{\alpha,\beta=x, y} D'_{\alpha\beta} q_\alpha q_\beta |\theta_{\bm{q} n}|^2, \label{eq:g1'} \\
    \mathrm{Tr}\bigl[(\mathcal{G} \Sigma)^2\bigr] &\approx \frac{1}{T} \sum_{\bm{q} n} \biggl[ -\chi \biggl( \nu_n \theta_{\bm{q} n} + \frac{\ii \phi_{\bm{q} n}}{\sqrt{N}} \biggr) \biggl( -\nu_n \theta_{\bm{q} n}^* + \frac{\ii \phi_{\bm{q} n}^*}{\sqrt{N}} \biggr) + \sum_{\alpha,\beta=x, y} D''_{\alpha\beta} q_\alpha q_\beta |\theta_{\bm{q} n}|^2 \biggr], \label{eq:g2'}
\end{align}
with
\begin{align}
    D'_{\alpha\beta} &= - \frac{1}{N} \sum_{\bm{k}} \frac{\xi_{\bm{k}}}{E_{\bm{k}}} \tanh\biggl(\frac{E_{\bm{k}}}{2T}\biggr) \frac{\partial^2 \xi_{\bm{k}}}{\partial k_\alpha \partial k_\beta}, \label{eq:D'} \\
    D''_{\alpha\beta} &= \frac{2}{N} \sum_{\bm{k}} n_F'(E_{\bm{k}}) \frac{\partial \xi_{\bm{k}}}{\partial k_\alpha} \frac{\partial \xi_{\bm{k}}}{\partial k_\beta}, \\
    \chi &= \frac{1}{N} \sum_{\bm{k}} \frac{\partial}{\partial \xi_{\bm{k}}} \biggl[ \frac{\xi_{\bm{k}}}{E_{\bm{k}}} \tanh\biggl(\frac{E_{\bm{k}}}{2T}\biggr) \biggr].
\end{align}
Here the frequency summations have been completed and $n_F(\cdot)$ is the Fermi distribution function and $n_F'(\cdot)$ represents its derivative.
We can integrate Eq.~(\ref{eq:D'}) by parts and use the periodicity of $\xi_{\bm{k}}$ to get the phase stiffness
\begin{equation}
    D_{\alpha\beta} \equiv D'_{\alpha\beta} + D''_{\alpha\beta} = \frac{1}{N} \sum_{\bm{k}} \frac{\partial}{\partial E_{\bm{k}}} \biggl[ \frac{1}{E_{\bm{k}}} \tanh\biggl(\frac{E_{\bm{k}}}{2T}\biggr) \biggr] \frac{1}{E_{\bm{k}}} \biggl( \frac{\xi_{\bm{k}}}{2} \frac{\partial |\Delta_{\bm{k}}|^2}{\partial k_\alpha} - |\Delta_{\bm{k}}|^2 \frac{\partial \xi_{\bm{k}}}{\partial k_\alpha} \biggr) \frac{\partial \xi_{\bm{k}}}{\partial k_\beta}. \label{eq:D}
\end{equation}
Note that both $D'$ and $D''$ are symmetric matrices, so is $D$.
From Eq.~(\ref{eq:D}), it is clear that $D=0$ if $\Delta_{\bm{k}}=0$ as it should.

Finally, plugging Eqs.~(\ref{eq:g1'}, \ref{eq:g2'}) into Eq.~(\ref{eq:Seff}) and integrating out the fields $\phi_{\bm{q}n}$ and $\theta_{\bm{q}n}$, we obtain the free energy as in Eq.~(\ref{eq:F}).

\end{widetext}

\end{document}